\documentclass{article}
\usepackage{iclr2027_conference,times}
\iclrfinalcopy
\usepackage{hyperref}
\usepackage{url}

\usepackage{amsmath,amssymb,amsthm,mathtools,bm,booktabs,microtype,xcolor}
\usepackage{graphicx}
\def\titlefull{BRACE: Blockwise Rank Aggregation for Correlation Estimation}

\newcommand{\E}{\mathbb E}
\newcommand{\Pp}{\mathbb P}
\newcommand{\Var}{\operatorname{var}}
\newcommand{\xiC}{\widehat\xi_{\rm CCC}}
\newcommand{\xiB}{\widehat\xi_{B,K}}

\newtheorem{theorem}{Theorem}
\newtheorem{proposition}[theorem]{Proposition}
\newtheorem{condition}[theorem]{Condition}

\author{
Man Hei Ngou$^{1}$, Yanran Li$^{2}$, Zhexiao Lin$^{3}$, Zexi Cai$^{1}$ \\
$^{1}$University of Macau \quad
$^{2}$Columbia University \quad
$^{3}$Two Sigma Investments \\
{\small\texttt{\{mc66392,zxcai\}@um.edu.mo \quad yl5465@columbia.edu \quad zhexiaolin@berkeley.edu}} }

\begin{document}
\title{\titlefull}
\maketitle

\begin{abstract}
Chatterjee's estimator uses only two local comparisons per interior observation, leaving a finite-replication variance gap under fixed alternatives. 
We introduce BRACE, short for blockwise rank aggregation for correlation estimation, which replaces adjacent comparisons with local blocks and averages every within-block rank difference. 
The block size controls the variance cost of finite local replication. 
An $L_2$ expansion separates the efficient first-order component of the response-rank statistic from an orthogonal finite-replication component. 
For an admissible diverging block size, the latter vanishes, so the direct rank estimator attains the information bound.
The decomposition yields a consistent variance estimator and Wald confidence intervals under fixed alternatives.
At independence, the efficient first-order term vanishes, and the proposed estimator enters a second-order regime in which the same factor controls its null variance.
\end{abstract}

\section{Introduction}
Chatterjee's correlation combines a general measure of functional dependence
with an exceptionally simple rank estimator
\citep{chatterjee2021,chatterjee2024survey}. For continuous $Y$,
\[
\xi(X,Y)=6\int \Var\{\Pp(Y\le t\mid X)\}\,dF_Y(t),
\]
so that $\xi=0$ characterizes independence and $\xi=1$ characterizes
functional dependence under the usual nondegeneracy conditions
\citep{dette2013,gamboa2018}. After ordering the sample by $X$, Chatterjee's
estimator uses only adjacent response ranks,
\[
\xiC=1-\frac{3}{n^2-1}\sum_{i=1}^{n-1}|R_{i+1}-R_i|.
\]
Chatterjee's estimator uses only two local comparisons per interior observation. 
For fixed alternatives satisfying the stated local regularity conditions, our decomposition identifies the resulting efficiency gap exactly: the entire asymptotic variance above the nonparametric efficiency bound is a canonical second-order component generated by finite local replication. 
We therefore introduce BRACE, short for blockwise rank aggregation for correlation estimation.
BRACE partitions the predictor-ordered sample into consecutive blocks and
averages all pairwise response-rank differences within each block.
A pair-specific decomposition separates the first-order influence function
from a canonical finite-replication term. Based on the decomposition, we show that,
for block size $K$, the BRACE estimator $\xiB$ satisfies
\[
n\Var(\xiB)
=\sigma_{\rm eff}^2+\frac{18\nu}{K-1}+o(1),
\]
where $\sigma_{\rm eff}^2$ is the nonparametric efficiency bound, 
and \(\nu\ge0\) measures the residual pairwise variation 
after the first-order influence is removed. 
The $L_2$ expansion of the response-rank statistic separates its efficient first-order term from an exactly orthogonal canonical edge term with variance $18\nu/\{n(K-1)\}+o(n^{-1})$. A slowly growing block size removes the latter while preserving rank-only $O(n\log n)$ computation, yielding a direct estimator that is asymptotically linear with the efficient influence function and attains the information bound.
At independence, the first-order term vanishes, 
and the same $K-1$ governs the exact null variance.

Several strands of rank-based dependence estimation are closely related to
BRACE.  The first develops the large-sample theory of the original
one-neighbour statistic.  \citet{gamboa2022} showed that the rank construction
estimates the Cram\'er--von Mises sensitivity index in the given-data setting
and extends to a broader class of global sensitivity indices.
For Chatterjee's estimator, \citet{linhanlimit} proved asymptotic normality
under fixed alternatives and constructed a consistent analytic variance
estimator, \citet{kroll2024} obtained asymptotic normality under weaker
distributional assumptions, and \citet{chhaibi2026} derived an explicit
fixed-alternative asymptotic variance.

A second strand studies the benefit of using more than one local comparison.
For independence testing, the one-neighbour statistic can have poor power
against smooth local alternatives \citep{shi2022,auddy2024}, while the
multiple-right-neighbour construction of \citet{linhan2023} can substantially
improve detection.  For a first-order Sobol functional,
\citet{kleinrochet2024} showed that averaging lagged rank estimators reduces
the additional variance of the lag-one estimator and can attain the efficiency
bound.  These results establish gains from local averaging under testing and
Sobol criteria, rather than the fixed-alternative estimation criterion studied
here.

A third strand develops more general graph-based dependence statistics.  The
kernel-and-graph framework of \citet{debghosalsen2020} includes
$k$-nearest-neighbour and minimum-spanning-tree constructions, and
\citet{azadkia2021} used nearest-neighbour graphs for multivariate and
conditional dependence measures.  Subsequent work studies intrinsic
dimension, rank invariance, bias control, and limit theory in more general
geometric settings
\citep{hanhuang2024,tranhan2024,azadkiachenhan2026,gaohanli2026}.
These methods accommodate multivariate geometry; in the scalar problem here,
the ordering of $X$ already determines the local geometry and permits a
specialized blockwise construction.

Most closely related to our efficiency question, \citet{klein2025efficiency}
derived the efficient influence function and information bound for the
corresponding Cram\'er--von Mises functional, together with an efficient
procedure based on estimating the conditional distribution of $Y$ given $X$.
What remained open was the structural source of the adjacent rank statistic's
efficiency gap and whether a direct rank construction could continuously
remove it.  BRACE identifies this gap with a canonical finite-replication
component and shows that balanced local replication removes it at the
inverse-replication rate.  The point estimator still requires only predictor
ordering and within-block response-rank comparisons; nuisance estimation is
needed only for feasible standard errors.

\section{Blockwise rank estimation}
\label{sec:method}
Write
\(
U=F_X(X), V=F_Y(Y).
\)
Under continuous margins, both $U,V\sim U(0,1)$. If $V'$ is conditionally
independent of $V$ given $U$ and has the same conditional law, then
\begin{equation}
\xi=1-3\E|V-V'|.
\label{eq:local-replicate-id}
\end{equation}
Hence, observations with nearby \(U\)-values can be treated as 
approximate conditional replicates. While \(\xiC\) uses only adjacent pairs, 
BRACE aggregates all pairwise comparisons within each local block.

Assume first that $n=mK$ for integers $m\ge1$ and $K\ge2$.
Order the observations by $X$ and, for notational convenience, relabel the
corresponding probability-scale pairs as $(U_i,V_i)$, $i=1,\ldots,n$, so that
$U_1<\cdots<U_n$ almost surely. Let $R_i$ denote the response rank attached to
the $i$th position in this predictor ordering. Define consecutive blocks
\[
B_b=\{(b-1)K+1,\ldots,bK\},\qquad b=1,\ldots,m,
\]
and let $M=m\binom{K}{2}=n(K-1)/2$ denote the total number of within-block
unordered pairs. We define the BRACE estimator by
\begin{equation}
\xiB
=1-\frac{3}{(n+1)M}
\sum_{b=1}^m\sum_{\substack{i<j,\,i,j\in B_b}}|R_i-R_j|.
\label{eq:brace}
\end{equation}
Each observation in a block is compared with exactly $K-1$ other observations. 
The fixed-alternative theory in Section~\ref{sec:fixed} shows that this structure
makes the finite-replication variance proportional to $1/(K-1)$.

The complete block sum does not require explicit enumeration of its
$\binom{K}{2}$ pairs. If the response ranks in block $b$ are sorted as
$r_{b,(1)}<\cdots<r_{b,(K)}$, then
\[
\sum_{1\le i<j\le K}|r_{b,(i)}-r_{b,(j)}|
=\sum_{j=1}^K(2j-K-1)r_{b,(j)}.
\]
Thus, the additional cost after the global ordering and response
ranking is $O(K\log K)$ per block, or $O(n\log K)$ in total. 
The full point estimator therefore remains $O(n\log n)$.

\section{Fixed-alternative efficiency and inference}
\label{sec:fixed}
To isolate the finite-replication variance, consider a block of \(K\) exact conditional replicates. Its pair average is an order-two U-statistic, whose canonical component contributes variance of order \(1/(K-1)\) after aggregation across blocks. Actual blocks contain observations with nearby, rather than identical, predictor values; under our regularity conditions, the resulting approximation error is negligible at the root-\(n\) scale. 
Write
\(
G_u(v)=\Pp(V\le v\mid U=u).
\)
For conditionally independent $Z_1,Z_2\sim G_u$, define
\[
g(u)=\E|Z_1-Z_2|,\qquad q(u,v)=\E|v-Z_2|,
\]
\[
r_u(v,w)=|v-w|-q(u,v)-q(u,w)+g(u),\qquad
\nu(u)=\E r_u(Z_1,Z_2)^2,
\]
and put $\nu=\E\nu(U)$. Consider $Z_1,\ldots,Z_K\stackrel{\rm iid}{\sim}G_u$. Its within-block U-statistic has the exact Hoeffding decomposition
\begin{equation}
\binom{K}{2}^{-1}\sum_{i<j}|Z_i-Z_j|
=g(u)+\frac2K\sum_{i=1}^K\{q(u,Z_i)-g(u)\}
+\frac{2}{K(K-1)}\sum_{i<j}r_u(Z_i,Z_j).
\label{eq:ideal-hoeffding}
\end{equation}
The last term is canonical and has variance
$2\nu(u)/\{K(K-1)\}$. Averaging $n/K$ blocks and applying the factor
$-3$ in \eqref{eq:local-replicate-id} gives the finite-replication
contribution $18\nu/\{n(K-1)\}$.

It remains to quantify the error from using nearby rather than identical
conditional laws. For arbitrary CDFs $G$ and $H$ on $[0,1]$, let
$Z,Z'\stackrel{\rm iid}{\sim}G$ and
$W,W'\stackrel{\rm iid}{\sim}H$, independently across the two pairs. Then
\begin{equation}
\E|Z-W|-\frac12\{\E|Z-Z'|+\E|W-W'|\}
=\int_0^1\{G(t)-H(t)\}^2\,dt.
\label{eq:cross-l2}
\end{equation}
For $u,u'\in[0,1]$, write
\[
d_2^2(u,u')=\int_0^1\{G_u(t)-G_{u'}(t)\}^2\,dt,
\qquad
d_\infty(u,u')=\sup_{t\in[0,1]}|G_u(t)-G_{u'}(t)|.
\]
Conditional on the ordered predictor values, the $V_i$ are independent 
with CDFs $G_{U_i}$. Applying \eqref{eq:cross-l2} within block $b$ gives
\begin{equation}
\E\left[
\binom{K}{2}^{-1}
\sum_{\substack{i<j,\,i,j\in B_b}}|V_i-V_j|
\,\middle|\, U_1,\ldots,U_n\right]
=\frac1K\sum_{i\in B_b}g(U_i)
+\binom{K}{2}^{-1}
\sum_{\substack{i<j,\,i,j\in B_b}}d_2^2(U_i,U_j).
\label{eq:block-localization}
\end{equation}
The first term is the block average of $g(U_i)$, 
and the second term is the exact conditional mean error 
induced by using nearby predictor values in place of identical conditional replicates.

Define
\[
B_{n,K}=\frac1M\sum_{b=1}^m
\sum_{\substack{i<j,\,i,j\in B_b}}d_2^2(U_i,U_j),
\qquad
\Delta_{n,K}=\frac1M\sum_{b=1}^m
\sum_{\substack{i<j,\,i,j\in B_b}}d_\infty^2(U_i,U_j).
\]
Here $B_{n,K}$ is the average second-order approximation term in the
conditional mean of \eqref{eq:block-localization}, whereas $\Delta_{n,K}$
controls the additional remainder from replacing the latent probability-scale
responses by empirical ranks. We impose the following regularity conditions.

\begin{condition}[Fixed-alternative regularity]
    Along the block-size sequence $K=K_n$, assume
    \begin{equation}
    K/n\to0,\qquad
    \E\Delta_{n,K}\to0,\qquad
    \sqrt n\,\{\E(B_{n,K}^2)\}^{1/2}\to0.
    \label{eq:local-average-main}
    \end{equation}
\end{condition}
To give a sufficient condition, consider the uniform H\"older bound
\[
\sup_{v\in[0,1]}|G_u(v)-G_{u'}(v)|\le C|u-u'|^s.
\]
Standard moments of uniform order-statistic spacings then give
\(
\E\Delta_{n,K}\lesssim(K/n)^{2s},
\)
and
\(
\{\E(B_{n,K}^2)\}^{1/2}\lesssim(K/n)^{2s}.
\)
Hence $\sqrt n(K/n)^{2s}\to0$ is sufficient, and
$K_n=\lceil\log n\rceil$ is admissible for every fixed $s>1/4$.

The semiparametric efficiency benchmark is adapted from \citet{klein2025efficiency} who derived the canonical gradient for the Cram\'er--von Mises functional underlying $\xi$. To give the efficiency bound, define
\(
\eta=\E\int_0^1G_U(t)^2\,dt.
\)
Since \(V\) is marginally uniform, \(\E G_U(t)=t\), and hence
\(
\xi =6\int_0^1\Var\{G_U(t)\}\,dt =6\eta-2.
\)
For \(u,v\in[0,1]\), define
\[
J(u,v)=\int_v^1G_u(t)\,dt,\quad
a(u)=\int_0^1G_u(t)^2\,dt,\quad
H(v)=\E\{G_U(v)^2\}.
\]
\begin{proposition}[Efficient influence function]
\label{thm:eif}
At every distribution with continuous margins,
\begin{equation}
\psi_\xi(U,V)
=
6\{2J(U,V)-a(U)+H(V)-2\eta\}
\label{eq:eif}
\end{equation}
is the canonical gradient of $\xi$ in the unrestricted dominated
nonparametric model, with efficiency bound
\(
\sigma_{\rm eff}^2
=
\E\{\psi_\xi(U,V)^2\}.
\)
\end{proposition}
To isolate the information benchmark against the finite-replication component,
we re-express the same gradient in terms of the conditional-replicate kernel.
When the conditional laws are atomless,
\begin{equation}
\psi_\xi(U,V)
=
-3\{2q(U,V)-g(U)-\theta+A(V)\},
\label{eq:eif-pairwise}
\end{equation}
where
\[
A(v)
=
2\E\!\left[
\{2G_U(V)-1\}\{1(v\le V)-V\}
\right],
\qquad
\theta=(1-\xi)/3.
\]
The BRACE expansion explains how this efficient first-order term emerges
from the actual empirical-rank statistic.  The first-order Hoeffding
projection of the local conditional-replicate kernel yields
$2q(U,V)-g(U)-\theta$.  Replacing the latent
$V_i=F_Y(Y_i)$ by $R_i/(n+1)$ introduces a second empirical interaction,
whose first-order Hoeffding projection is $A(V_i)$.  The two first-order
terms combine to recover $\psi_\xi$, while the remaining canonical
within-block component is orthogonal to all one-observation terms and has
variance quantified in Theorem~\ref{thm:fixed}.

\begin{theorem}[BRACE variance continuum]
\label{thm:fixed}
Assume the conditional laws are atomless and
\eqref{eq:local-average-main} holds along $K=K_n$. If
$K_n\to K_0\in\{2,3,\ldots\}$, then
\[
\sqrt n\{\widehat\xi_{B,K_n}-\xi\}
\Rightarrow
N\!\left(0,\sigma_{\rm eff}^2+\frac{18\nu}{K_0-1}\right).
\]
If $K_n\to\infty$, then
\[
\sqrt n\{\widehat\xi_{B,K_n}-\xi\}
\Rightarrow N(0,\sigma_{\rm eff}^2),
\]
so the limiting variance coincides with the nonparametric efficiency bound.
More generally,
\begin{equation}
n\Var(\widehat\xi_{B,K_n})
=\sigma_{\rm eff}^2+\frac{18\nu}{K_n-1}+o(1).
\label{eq:variance-continuum}
\end{equation}
\end{theorem}
The expansion gives an $L_2$ decomposition: 
the efficient influence function is accompanied by an orthogonal canonical
within-block component with variance
\(18\nu/{n(K-1)}+o(n^{-1})\).
For admissible \(K_n\to\infty\), this canonical component vanishes at the
root-\(n\) scale, so BRACE is asymptotically linear with influence function
\eqref{eq:eif} and attains the nonparametric efficiency bound. 
Theorem~\ref{thm:fixed} also yields a direct comparison with the original
\(\xiC\), whose fixed-alternative asymptotic normality has been established
in \citet{linhanlimit,kroll2024}.

\begin{proposition}[Comparison with Chatterjee's estimator]
Under the fixed-alternative conditions stated above, we have
\[
n\Var(\xiC)=\sigma_{\rm eff}^2+9\nu+o(1).
\]
Therefore, BRACE with $K=3$ matches its first-order variance, and every fixed $K>3$ improves it
whenever $\nu>0$.
\end{proposition}
\subsection{Variance estimation under fixed alternatives}
The asymptotic variance of BRACE has two components. We estimate the 
efficiency-bound component by the empirical second moment of estimated canonical gradients, 
and estimate the canonical variance \(\nu\) separately. 
The BRACE point estimator uses only sorting and response ranks. Feasible standard errors estimate the efficient first-order component by cross-fitting and the canonical component $\nu$ by local rank moments.
Denote the resulting estimators by \(\widehat\sigma_{\rm eff}^2\) and \(\widehat\nu\), 
and define
\(
\widehat V_K
=\widehat\sigma_{\rm eff}^2
+\frac{18}{K-1}\widehat\nu.
\)
\begin{theorem}[Feasible variance estimation]
\label{thm:variance-estimation}
Under Theorem~\ref{thm:fixed} and standard nuisance-consistency conditions,
$\widehat V_K\to_p\sigma_{\rm eff}^2+18\nu/(K-1)$ for fixed $K$, while
$\widehat V_{K_n}\to_p\sigma_{\rm eff}^2$ for admissible $K_n\to\infty$.
If $\sigma_{\rm eff}^2>0$, the corresponding Wald interval has asymptotic
coverage $1-\alpha$.
\end{theorem}

\section{Independence regime}
\label{sec:independence}
At independence, $G_u(v)=v$ and the influence function
\eqref{eq:eif} vanishes. 
In our formulation, the fixed-alternative first-order expansion 
gives way to a canonical second-order limit.
\begin{theorem}[Exact null variance and null CLT]
\label{thm:null}
If $X$ and $Y$ are independent with continuous margins and $n=mK$, then
\begin{equation}
\E_0(\xiB)=0,\qquad
\Var_0(\xiB)
=\frac{4(n-K)}{5n(K-1)(n+1)}.
\label{eq:null-exact}
\end{equation}
If $K_n/n\to0$, then
\[
\sqrt{n(K_n-1)}\,\widehat\xi_{B,K_n}\Rightarrow N(0,4/5).
\]
For fixed $K$, this is equivalent to
\[
\sqrt n\,\xiB\Rightarrow
N\{0,4/[5(K-1)]\}.
\]
In particular, $K = 3$ again recovers the variance 2/5 of Chatterjee's statistic.
\end{theorem}
For fixed \(K\), the role of the block size can be seen directly by comparing the 
fixed-alternative and null variances. Let \(\operatorname{avar}(\widehat\xi)\) denote 
the limit of \(n\operatorname{Var}(\widehat\xi)\) under a fixed alternative, and \(\operatorname{avar}_0(\widehat\xi)\) the corresponding limit under independence. 
Whenever $\nu>0$,
\begin{equation}
\frac{\operatorname{avar}(\xiB)-\sigma_{\rm eff}^2}
{\operatorname{avar}(\xiC)-\sigma_{\rm eff}^2}
=\frac{\operatorname{avar}_0(\xiB)}
{\operatorname{avar}_0(\xiC)}
=\frac2{K-1}.
\label{eq:replication-factor-null}
\end{equation}
Therefore, the same factor \(2/(K-1)\) governs two distinct asymptotic regimes. 
Under fixed alternatives, it is the ratio of the excess variances 
above the efficiency bound, whereas under independence, it is the ratio of the 
full asymptotic variances.

The factor \(2/(K-1)\) can also be understood from the number of 
local pairwise comparisons involving each observation. 
In BRACE, every observation is compared with the other \(K-1\) observations in its block. 
In Chatterjee's statistic, an interior observation enters two adjacent rank differences. 
Hence, the relevant comparison counts are \(K-1\) for BRACE and \(2\) for Chatterjee's statistic. 
Correspondingly, the canonical finite-replication variance terms are \(18\nu/(K-1)\) 
and \(18\nu/2=9\nu\), respectively.

\section{Numerical validation}
\label{sec:numerical}
The numerical study examines two results of the preceding theory, i.e., the
finite-$K$ variance continuum in \eqref{eq:variance-continuum}, and the
fixed-alternative inference procedure in Theorem~\ref{thm:variance-estimation}. 
We use the Gaussian rotation model
\[
X\sim N(0,1),\qquad
Y=\rho X+(1-\rho^2)^{1/2}\varepsilon,\qquad
\varepsilon\sim N(0,1),
\]
for which
\[
\xi(\rho)=-\frac12+\frac3\pi
\arcsin\!\left(\frac{1+\rho^2}{2}\right).
\]
For this model,
$K_n=\lceil\log n\rceil$ is admissible for every fixed $0<|\rho|<1$.
The population variance components can also be evaluated numerically. 
For $\rho=0.3$, $(\sigma_{\rm eff}^2,\nu)=(0.10579,0.03947)$, 
while for $\rho=0.6$ they are $(0.27689,0.02608)$. 

Table~\ref{tab:variance-continuum} compares the empirical $n$-scaled variance
at $n=1000$ with the fixed-$K$ prediction
$\sigma_{\rm eff}^2+18\nu/(K-1)$. Chatterjee's estimator and 
the $K=\infty$ efficiency bound are included as benchmarks. 
At $K=3$, the theoretical variance coincides with that of Chatterjee's estimator, 
and the empirical variances are also close. Increasing $K$ from 3 to 5 and 10 lowers
the empirical variance in the amount predicted by the variance formula, 
and shows a converging trend towards the efficiency bound.

\begin{table}[!htpb]
\centering
\caption{Empirical and theoretical $n$-scaled variances at $n=1000$, based on
5000 Monte Carlo replications. The $K=\infty$ entries are efficiency bounds.
The largest Monte Carlo s.e. of an empirical $n\Var$ is 0.010.}
\label{tab:variance-continuum}
\resizebox{0.9\textwidth}{!}{%
\begin{tabular}{cccccccc}
\toprule
\multicolumn{4}{c}{$\rho=0.3$}&\multicolumn{4}{c}{$\rho=0.6$}\\
estimator & $K$ & empirical $n\Var$ & theory $n\Var$
& estimator & $K$ & empirical $n\Var$ & theory $n\Var$\\
\midrule 
$\xiC$ & -- & 0.454 & 0.461
& $\xiC$ & -- & 0.500 & 0.512 \\
$\xiB$ & 3 & 0.460 & 0.461
& $\xiB$ & 3 & 0.508 & 0.512 \\
$\xiB$ & 5 & 0.283 & 0.283
& $\xiB$ & 5 & 0.385 & 0.394 \\
$\xiB$ & 10 & 0.182 & 0.185
& $\xiB$ & 10 & 0.325 & 0.329 \\
-- & $\infty$ & -- & 0.106
& -- & $\infty$ & -- & 0.277 \\
\bottomrule 
\end{tabular}
}
\end{table}

We next assess feasible variance
estimation, taking $K=\lceil\log n\rceil$ throughout. 
For each $n\in\{500,1000,2000,4000\}$ and $\rho\in\{0.3,0.6\}$, 
Table~\ref{tab:variance-inference} reports the empirical
$n$-scaled variance, the theoretical value of
$\sigma_{\rm eff}^2+18\nu/(K-1)$, the mean estimates of the two
variance components, the mean feasible variance estimate, and the associated
95\% Wald coverage, based on 5000 Monte Carlo replications.

The empirical $n$-scaled variances agree closely with the theoretical values,
and the variance estimator also tracks the sampling variance.  The estimates
of $\nu$ are stable across sample sizes, while the cross-fitted estimate of
$\sigma_{\rm eff}^2$ shows modest finite-window bias, more visibly at
$\rho=0.6$ for the smaller sample sizes.  The empirical coverages remain
close to the nominal level.
\begin{table}[!htpb]
\centering
\caption{Variance-component estimation and inference for
$K=\lceil\log n\rceil$, based on 5000 Monte Carlo replications per setting.
The largest Monte Carlo s.e.s are 0.0073 for empirical $n\Var$, 0.0004 for
$E\widehat\sigma_{\rm eff}^2$, 0.003 for $100E\widehat\nu$, 0.0004 for
$E\widehat V_K$, and 0.0034 for coverage.}
\label{tab:variance-inference}
\resizebox{0.8\textwidth}{!}{%
\begin{tabular}{cccccccc}
\toprule
$(n,K)$ & $\rho$ & empirical $n\Var$ & theory $n\Var$
& $E\widehat\sigma_{\rm eff}^2$ & $100E\widehat\nu$
& $E\widehat V_K$ & coverage \\
\midrule 
$(500,7)$  & 0.3 & 0.227 & 0.224 & 0.106 & 3.95 & 0.224 & 0.952 \\
$(500,7)$  & 0.6 & 0.359 & 0.355 & 0.258 & 2.65 & 0.337 & 0.942 \\
$(1000,7)$ & 0.3 & 0.219 & 0.224 & 0.103 & 3.95 & 0.222 & 0.947 \\
$(1000,7)$ & 0.6 & 0.355 & 0.355 & 0.262 & 2.63 & 0.341 & 0.941 \\
$(2000,8)$ & 0.3 & 0.203 & 0.207 & 0.102 & 3.95 & 0.204 & 0.955 \\
$(2000,8)$ & 0.6 & 0.342 & 0.344 & 0.266 & 2.63 & 0.333 & 0.942 \\
$(4000,9)$ & 0.3 & 0.190 & 0.195 & 0.102 & 3.95 & 0.191 & 0.950 \\
$(4000,9)$ & 0.6 & 0.328 & 0.336 & 0.268 & 2.62 & 0.327 & 0.951 \\
\bottomrule
\end{tabular}
}
\end{table}

\section{Real-data analysis}
\label{sec:realdata}

We illustrate fixed-alternative estimation and inference using the Combined
Cycle Power Plant data from the UCI Machine Learning Repository
\citep{tufekci2014ccpp}.  The data contain 9568 hourly-average records
collected during full-load operation between 2006 and 2011.  The response is
net electrical energy output (PE), and the environmental predictors are
ambient temperature (AT), exhaust vacuum (V), ambient pressure (AP), and
relative humidity (RH).

The released measurements are rounded and therefore contain ties.  For a
recorded variable $W$ with distribution function $F_W$, let $A_W\sim U(0,1)$
be independent of the data and define the distributional transform
\[
W^\dagger=F_W(W^-)+A_W\{F_W(W)-F_W(W^-)\}.
\]
This transform is uniform even when $F_W$ has atoms \citep{ruschendorf2009}.
We generate independent auxiliary keys for predictor and response ties once
and retain the resulting strict orderings throughout the analysis.  Thus the
population target for a predictor $X$ is
\[
\xi^\dagger(P_{\rm CCPP})=\xi(X^\dagger,\mathrm{PE}^\dagger),
\]
with the auxiliary uniforms integrated into the functional.  The point and
variance estimators use the same resolved orderings.

AT, AP, and RH have 2773, 2517, and 4546 distinct observed values,
respectively, whereas V has only 634.  We focus the precision comparison on
AT, AP, and RH because the much coarser support of V makes its estimate
sensitive to the localization scale.  We model the released records as iid
draws from a distribution of full-load operating records and use the default
block size $K=\lceil\log n\rceil=10$.  Wald intervals use the variance
estimator in Theorem~\ref{thm:variance-estimation} under the fixed-alternative
regularity condition.

To summarize the estimated precision difference, define the standard-error ratio
\[
\frac{\widehat{\mathrm{SE}}_C}{\widehat{\mathrm{SE}}_B}
=\left\{
\frac{\widehat\sigma_{\rm eff}^2+9\widehat\nu}
     {\widehat\sigma_{\rm eff}^2+18\widehat\nu/(K-1)}
\right\}^{1/2}.
\]
The two estimators give similar point estimates, while
BRACE has smaller estimated standard errors for all three predictors.  The
gain is largest for RH, where the estimated standard-error ratio is $1.43$.

\begin{table}[!htpb]
\centering
\caption{Combined Cycle Power Plant data. BRACE uses $K=10$. The intervals
are 95\% Wald intervals under the iid sampling model and the regularity
conditions in the text.}
\label{tab:ccpp}
\begin{tabular}{ccccc}
\toprule
predictor & $\xiC$ & $\xiB$ & $\xiB$ 95\% CI &
$\widehat{\mathrm{SE}}_C/\widehat{\mathrm{SE}}_B$ \\
\midrule
AT & 0.7041 & 0.7039 & $[0.6979,\,0.7100]$ & 1.13 \\
AP & 0.2062 & 0.2107 & $[0.1997,\,0.2218]$ & 1.28 \\
RH & 0.1087 & 0.1007 & $[0.0909,\,0.1104]$ & 1.43 \\
\bottomrule
\end{tabular}
\end{table}

\section{Discussion}
\label{sec:discussion}
BRACE separates the ordered-rank error into first-order information, finite local replication, and block approximation. Under fixed alternatives, the first component gives the efficiency bound, while finite replication contributes the orthogonal excess variance $18\nu/(K-1)$; the block-approximation error is governed by squared $L_2$ discrepancies between neighbouring conditional laws. The same inverse-degree factor $2/(K-1)$ governs the asymptotic null-variance ratio relative to Chatterjee's statistic. Data-adaptive choice of $K$ for second-order finite-sample risk remains open.

The present theory assumes continuous margins, a scalar ordering variable, and independent
observations. Discrete margins, multivariate predictors, and dependent data require different
approximation error and variance analysis. 
Uniform inference across the transition from the fixed alternative to the null is a separate problem.

\bibliographystyle{iclr2027_conference}
\bibliography{references}
\end{document}